\documentclass[sigconf]{acmart}

\usepackage{multicol}
\usepackage{multirow}
\usepackage{adjustbox}
\usepackage{enumitem}
\usepackage[ruled,linesnumbered]{algorithm2e}
\usepackage{pifont}
\usepackage{caption,subcaption}
\usepackage{colortbl} 

\graphicspath{{figures/}}

\AtBeginDocument{%
  }
    
\copyrightyear{2025}
\acmYear{2025}
\setcopyright{acmlicensed}
\acmConference[KDD '25] {Proceedings of the 31st ACM SIGKDD Conference on Knowledge Discovery and Data Mining V.2}{August 3--7, 2025}{Toronto, ON, Canada.}
\acmBooktitle{Proceedings of the 31st ACM SIGKDD Conference on Knowledge Discovery and Data Mining V.2 (KDD '25), August 3--7, 2025, Toronto, ON, Canada}
\acmISBN{979-8-4007-1454-2/25/08}
\acmDOI{10.1145/3711896.3737127}

\acmSubmissionID{rtfb1630}

\begin{document}

\title{Shapley Value-driven Data Pruning for Recommender Systems}

\author{Yansen Zhang}
\orcid{0000-0001-8426-8837}
\affiliation{%
  \institution{City University of Hong Kong}
  \city{Hong Kong SAR}
  \country{China}
}
\email{yanszhang7-c@my.cityu.edu.hk}

\author{Xiaokun Zhang}
\authornote{Corresponding authors.}
\orcid{0000-0002-9755-2471}
\affiliation{%
  \institution{City University of Hong Kong}
  \city{Hong Kong SAR}
  \country{China}
}
\email{dawnkun1993@gmail.com}

\author{Ziqiang Cui}
\orcid{0000-0002-1742-7866}
\affiliation{%
  \institution{City University of Hong Kong}
  \city{Hong Kong SAR}
  \country{China}
}
\email{ziqiang.cui@my.cityu.edu.hk}

\author{Chen Ma}
\authornotemark[1]
\orcid{0000-0001-7933-9813}
\affiliation{%
  \institution{City University of Hong Kong}
  \city{Hong Kong SAR}
  \country{China}
}
\email{chenma@cityu.edu.hk}

\renewcommand{\shortauthors}{Yansen Zhang, Xiaokun Zhang, Ziqiang Cui, and Chen Ma}

\begin{abstract}


Recommender systems often suffer from noisy interactions like accidental clicks or popularity bias. Existing denoising methods typically identify users' intent in their interactions, and filter out noisy interactions that deviate from the assumed intent. However, they ignore that interactions deemed noisy could still aid model training, while some ``clean'' interactions offer little learning value. To bridge this gap, we propose Shapley Value-driven Valuation (SVV), a framework that evaluates interactions based on their objective impact on model training rather than subjective intent assumptions. In SVV, a real-time Shapley value estimation method is devised to quantify each interaction's value based on its contribution to reducing training loss. Afterward, SVV highlights the interactions with high values while downplaying low ones to achieve effective data pruning for recommender systems. In addition, we develop a simulated noise protocol to examine the performance of various denoising approaches systematically. Experiments on four real-world datasets show that SVV outperforms existing denoising methods in both accuracy and robustness. Further analysis also demonstrates that our SVV can preserve training-critical interactions and offer interpretable noise assessment. This work shifts denoising from heuristic filtering to principled, model-driven interaction valuation.

\end{abstract}





\maketitle

\newcommand\kddavailabilityurl{https://doi.org/10.5281/zenodo.15487805}

\ifdefempty{\kddavailabilityurl}{}{
\begingroup\small\noindent\raggedright\textbf{KDD Availability Link:}\\
The source code of this paper has been made publicly available at \url{\kddavailabilityurl}.
\endgroup
}

\section{Introduction}
\label{sec:intro}



Recommender systems are prone to noisy interactions~\cite{lee2021bootstrapping,wang2021denoising,wang2022learning,huang2023negative}, including popularity bias, accidental clicks, and spurious correlations induced by promotions or social trends. Such noisy interactions can severely degrade recommendation performance. For instance, an accidental click driven by exposure bias may lead the model to overestimate interest in irrelevant items, while purchases influenced by external factors can introduce misleading correlations that hinder effective training. These challenges introduce uncertainty in user preference modeling, making it difficult to distinguish informative interactions from noisy ones. Consequently, identifying the high-quality interactions in recommender systems is essential for improving recommendation accuracy and robustness.

To address this issue, various denoising approaches have been proposed, including auxiliary signal-based filtering~\cite{fox2005evaluating,kim2014modeling,bian2021denoising,lu2018between}, sample dropping~\cite{ding2019sampler,lin2023autodenoise,wang2021denoising,he2024double}, sample reweighting~\cite{gantner2012personalized,wang2021denoising,wang2023efficient,ge2023automated}, and robust learning~\cite{wu2016collaborative,wang2022learning,gao2022self}. Early methods leverage additional behavioral signals or item-side information to identify noisy interactions but often rely on manual labeling or extra feedback, limiting their scalability.  More recent strategies remove noisy interactions by dropping those with high loss (e.g., T-CE~\cite{wang2021denoising}) or reweighting them based on loss metrics (e.g., R-CE~\cite{wang2021denoising}, AutoDenoise~\cite{ge2023automated}), and autoencoder-based approaches (e.g., CDAE~\cite{wu2016collaborative}) attempt to reconstruct the input to differentiate clean from noisy data.

However, existing methods often assume that deviations from expected user behavior indicate low-quality interactions, equating behavioral intent with training utility. This assumption is problematic because interactions labeled as noisy, such as serendipitous clicks, may still encode valuable latent patterns, while interactions deemed ``clean'', such as manipulative purchases, might be misleading. Furthermore, these approaches typically rely on heuristic rules or opaque black-box models to define noise, lacking a principled mechanism to objectively evaluate the impact of individual interactions on the model's training. Such limitations hinder interpretability and risk removing interactions critical to training convergence, ultimately limiting recommendation robustness.

\begin{figure}[t]
    \centering
    \includegraphics[width=\linewidth]{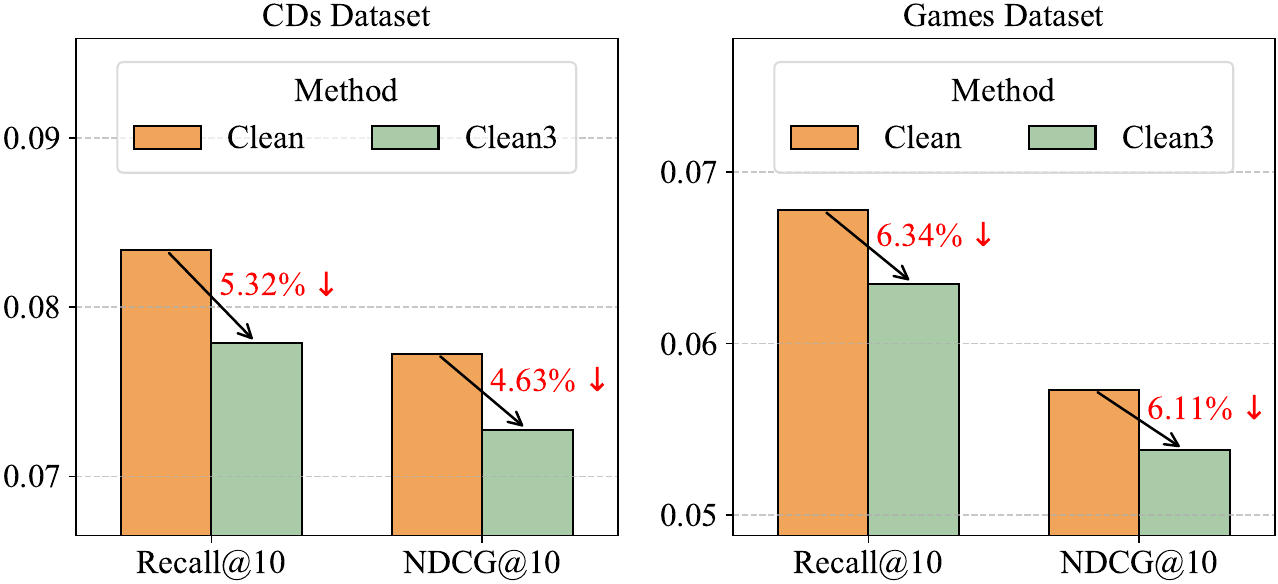}
    \caption{Performance comparisons of the Clean (all rating score ([1, 5])) training and Clean3 (all rating score ([1, 5]) $\geq$ 3) training in terms of Recall@10 and NDCG@10 over DAE.}
    \Description{Performance comparisons of the Clean (all rating score ([1, 5])) training and Clean3 (all rating score ([1, 5]) $\geq$ 3) training in terms of Recall@10 and NDCG@10 over DAE.}
    \label{fig:intro}
\end{figure}

To further interrogate these limitations, we conducted a controlled experiment on two real-world datasets using explicit rating data as a proxy for approximate ground-truth noise; the results are shown in Figure~\ref{fig:intro}. In our setup, each explicit rating was treated as an interaction, with ratings below 3 labeled as noises following common practice~\cite{wang2021denoising,he2020lightgcn}. Using a denoising autoencoder (DAE) as our baseline\footnote{Refer to Sec.~\ref{sec:basemodel} for more details about the baseline setting.}, training solely on ``high-quality'' interactions (Clean3) resulted in a 5.32\% and 4.63\% decline in Recall@10 and NDCG@10 on \textit{CDs}, and a 6.34\% and 6.11\% decline in Recall@10 and NDCG@10 on \textit{Gamses} compared to training on the full dataset (Clean) that includes ``noise''. These findings challenge the conventional wisdom that behavioral noise (e.g., low ratings) inherently harms model training. Instead, such interactions may carry valuable latent signals, while seemingly ``clean'' interactions may be less informative or even misleading. Therefore, denoising should focus on training utility rather than relying solely on subjective intent.


To develop an interpretable data pruning method that effectively distinguishes high-quality interactions, a key challenge is the ground-truth noise is often unavailable. This absence of a reliable benchmark makes it difficult to verify whether interactions are truly detrimental to model learning. To address this, we introduce a simulated noise\footnote{Noise here refers to low-quality interaction data, and in the following, we use noise and low-quality interchangeably.} injection protocol inspired by data debugging principles~\cite{karlavs2022data,grafberger2022data} to validate the impact of noise and establish a verifiable ground truth. Training on this noise-corrupted dataset\footnote{The benchmark generation process is detailed in Sec.~\ref{sec:benchmark}, and experimental comparisons can be found in Table~\ref{tab:noise-comparisons}.} leads to significant performance drops compared to the Clean setting. These results confirm the harmful influence of the injected noise and, critically, provide a verifiable ground truth: the known noise markers enable objective evaluation of denoising methods.

In this work, we propose \textbf{Shapley Value-driven Valuation (SVV)}, a framework that evaluates interactions based on their objective impact on model training rather than subjective intent assumptions. SVV overcomes the limitations of intent-centric denoising by directly evaluating interactions based on their training utility, defined as their objective impact on model learning. Our key innovation lies in quantifying an interaction's value by its marginal contribution to reducing training loss, measured by how the removal of a user--item interaction affects overall loss minimization. Specifically, we first train a base model (DAE) and then compute Shapley values for all observed interactions using FastSHAP~\cite{jethani2022fastshap}, a real-time estimation method that mitigates the high computational cost of exact Shapley value calculation.
This approach inherently captures the collaborative context, as interactions are valued based on their synergistic effects rather than in isolation. For example, the contribution of a low-rated item depends on its role in refining latent user and item representations. Crucially, SVV operates without manual heuristics or external data; the ``value'' scores of the interactions are derived solely from the model's training process. To validate the reliability, we inject simulated noise and demonstrate that interactions with low Shapley values align with these injected markers, thereby establishing an objective benchmark for denoising accuracy. By prioritizing training utility over subjective intent assumptions, SVV offers a transparent and theoretically grounded alternative to black-box denoising strategies, ensuring robust and interpretable data pruning.

To summarize, the main contributions of this work are as follows:
\begin{itemize}[leftmargin=*]
    \item We find that ``noise'' data determined by user intent can contribute to model training. Thus, this work introduces data pruning by evaluating its impact on model performance. To our knowledge, this is the first attempt to prune data from a model-driven interaction valuation perspective in recommender systems.
    \item We propose Shapley Value-driven Valuation (SVV), a principled framework that quantifies interaction importance through their context-aware and theoretically grounded contributions to training loss reduction.
    \item Extensive experiments on four real-world datasets validate the effectiveness, applicability, and interpretability of our method.
\end{itemize}

\section{Related Work}

\subsection{Denoising in Recommendation}

Recommender systems play an important role in our real lives and have attracted a lot of attention. However, recent studies~\cite{lee2021bootstrapping,wang2021denoising,huang2023negative} highlight that the recommender systems are often contaminated by factors like popularity bias~\cite{chen2023bias}, introducing substantial noise that degrades recommendation performance~\cite{wang2021denoising,wang2022learning}. To mitigate this issue, various approaches have been proposed to denoise implicit feedback. Early efforts leverage auxiliary user behaviors (e.g., `skip'~\cite{fox2005evaluating}, `dwell time'~\cite{kim2014modeling}, `like'~\cite{bian2021denoising}) and item side information~\cite{lu2018between} to identify noisy data. However, these methods require additional feedback and manual labeling, limiting their practicality. More recent approaches aim to denoise implicit feedback without external signals, which can be broadly categorized into sample dropping~\cite{ding2019sampler,lin2023autodenoise,wang2021denoising,he2024double}, sample reweighting~\cite{gantner2012personalized,wang2021denoising,wang2023efficient,ge2023automated}, and robust learning methods~\cite{wu2016collaborative,wang2022learning,gao2022self}. Drop-based methods filter out noisy interactions before training. Identifying clean samples is critical to their effectiveness. For instance, T-CE~\cite{wang2021denoising} empirically observes that noisy samples exhibit higher loss values, using this property for effective filtering. AutoDenoise~\cite{lin2023autodenoise} models denoising as a search-and-decision process, employing reinforcement learning for automation. Reweighting-based methods reduce the influence of noisy samples on model updates. R-CE~\cite{wang2021denoising} assigns lower weights to high-loss samples, while AutoDenoise~\cite{ge2023automated} integrates loss signals with user and item representations for weight computation. BOD~\cite{wang2023efficient} formulates weight learning as a bi-level optimization problem, achieving efficient and effective parameter estimation. Additionally, robust recommendation methods incorporate denoising strategies. Autoencoder-based models, such as CDAE~\cite{wu2016collaborative}, corrupt user interactions with random noise and attempt to reconstruct the original signals~\cite{sedhain2015autorec,wu2016collaborative}. DeCA~\cite{wang2022learning} leverages predictive divergence between models to distinguish clean from noisy samples. SGDL~\cite{gao2022self} observes the memorization effect of noise and removes noisy interactions during pretraining. Beyond general approaches, denoising has also been explored in specific contexts. In graph-based collaborative filtering, methods like RocSE~\cite{ye2023towards} remove noisy interaction edges. Domain-specific techniques target movie~\cite{zhao2023uncovering}, social recommendations~\cite{wang2023denoised} and so on, while sequential recommendation methods~\cite{zhang2021self,chen2022denoising} refine historical interactions and diversified recommendation~\cite{zhang2025cmb} detects the meaningful factors that affect the recommendation diversity.

Crucially, none of these methods provides a principled framework to quantify why specific interactions degrade performance or how much they contribute to model training--a gap our work addresses through game-theoretic data valuation.

\subsection{Data Valuation}

Data valuation quantifies the contribution of individual data points to algorithmic predictions and decisions, recognizing data as a fundamental driver of data science. A traditional approach is the leave-one-out (LOO) test~\cite{Cook00}, which evaluates the impact of removing a single sample on model outputs.
However, LOO ignores interactions between data points and performs poorly in many real-world applications~\cite{datashapley,chen2023algorithms}. To address these limitations, Shapley value-based methods, originating from cooperative game theory~\cite{shapley1953value}, have been proposed~\cite{JiaDWHGLZSS19, datashapley, yan2021if, chen2023algorithms}. Shapley values provide an intuitive and theoretically grounded measure of each data point's contribution, making them widely adopted in recent studies. However, their exact computation is computationally expensive, limiting practical applications. To mitigate this challenge, several acceleration techniques have been developed~\cite{jethani2022fastshap, chen2023harsanyinet}. Due to their strong interpretability and theoretical foundation, Shapley value-based methods have primarily been applied to traditional regression and classification tasks for feature importance analysis. A notable application domain is anomaly detection~\cite{ma2021comprehensive, li2023survey}. For instance, some studies~\cite{sundararajan2020many} compute Shapley values of reconstruction errors in PCA-based anomaly detection, while others~\cite{jakubowski2021anomaly} use Shapley values to assess feature contributions in autoencoder-based models. In recommender systems, Shapley value has also been explored for data valuation and interpretability, including enhancing data efficiency via reinforcement learning~\cite{jia2025beyond} and quantifying triplet importance for personalized ranking~\cite{he2024interpretable}.

Our work bridges this gap by adapting Shapley value-based valuation to user--item interaction data, quantifying contributions while accounting for contextual recommendation noise such as exposure bias and accidental clicks. This introduces a theoretically grounded framework for examining data quality in recommender training pipelines by focusing on interaction-level contributions.





\section{Preliminaries}
\label{sec:preliminary}

\subsection{Problem Formulation}

Due to the large volumes and accessibility of implicit feedback, this work focuses on recommender systems based on implicit user interactions. Given a user set $\mathcal{U}$ with $M$ users and an item set $\mathcal{V}$ with $N$ items, we define the user--item interaction matrix $\textbf{R} \in \mathbb{R}^{M \times N}$ based on users' implicit feedback:
\begin{equation}
    R_{u,v} =
    \begin{cases}
    1, & \text{if user $u$ interacted with item $v$;}\\
    0, & \text{otherwise.}
    \end{cases}
\end{equation}
We denote the user--item interactions as $D = \{(u, v) \mid R_{u,v}\}$, and the subset of observed interactions is denoted as $D_+ = \{(u, v) \mid R_{u,v} = 1\}$, which form the basis for training the recommendation models.
The goal of implicit feedback recommendation is to estimate scores for unobserved entries in $\textbf{R}$ and rank items for each user.

In this context, we aim to remove the noise interactions or select a high-quality subset $S \subseteq D_+$ from the observed interactions to train the recommendation model. The objective is to identify the subset $S$ that maximizes the model’s performance, defined as:
\begin{equation}
    S^* = \arg\max_{S \subseteq D_+} \mathcal{P}(f(S)),
\end{equation}
where $\mathcal{P}$ represents the model's performance, like accuracy, and $f(S)$ is the recommendation model trained on the selected high-quality subset $S$. The challenge lies in the absence of explicit quality labels for interactions, necessitating effective methods to discern and retain user--item interactions that enhance the model's predictive accuracy while filtering out low-quality data.

\subsection{Shapley Values}

Shapley values are a concept in cooperative game theory, assessing the contribution of each player to the total payoff of a coalition~\cite{shapley1953value}. For any cooperative game (e.g., value function in machine learning) $v: 2^I \mapsto \mathbb{R}$, where $I\equiv\{1, \cdots, |I|\}$ represents all players (features), the Shapley values $\phi_i(v) \in \mathbb{R}$ assigned to each feature $i = 1, \cdots, |I|$, are given by the formula:
\begin{equation}
    \phi_i(v) = {\frac{1}{|I|} \sum_{S \subseteq I \setminus \{i\}} {\binom{|I|-1}{|S|}}^{-1}} \Big(v(S \cup \{i\}) - {v(S)}\Big).
    \label{shapley_value}
\end{equation}
The difference $v(S \cup \{i\}) - v(S)$ represents the $i$-th feature's marginal contribution to the subset $S$, and the summation represents a weighted average across all subsets that do not include $i$.
In the model explanation context, the value function is chosen to represent how an individual prediction varies as different subsets of features are removed.
For example, given a prediction model $\mathbf{y} = f(\mathbf{x})$ that can handle the input with subset features (i.e., $\mathbf{x}_S$), and an input-output pair $(\mathbf{x}, \mathbf{y})$, the value function $v_{\mathbf{x},\mathbf{y}}$ can be represented as:
\begin{equation}
    v_{\mathbf{x},\mathbf{y}}(\mathbf{S}) = g \left(f(\mathbf{x}_S); \mathbf{y}  \right),
\end{equation}
where $\mathbf{S} \in \mathbb{R}^d$ represents the multi-hot vector consisting of the features involved in the input, and the function $g(\cdot)$ corresponds to the model behavior that a method is designed to explain.


\section{Methodology}
\label{sec:method}

\subsection{Base Recommendation Model}
\label{sec:basemodel}
In this work, we aim to identify high-quality user--item interactions by quantifying the marginal contribution of each item to overall model performance using Shapley values. To achieve this, we first reformulate the problem by treating items as features. Specifically, for a given interaction matrix $\mathbf{R}$, each user $u \in \mathcal{U} = \{1, \cdots, M\}$ is represented by a multi-hot vector $\mathbf{r}_u = (R_{u,1}, \cdots, R_{u,N}) \in \mathbb{R}^N$, where each element indicates whether a user has interacted with a corresponding item. This representation allows us to compute the importance of each interaction in a precise, feature-based manner and subsequently remove those interactions flexibly.

A key requirement for calculating Shapley values is that the base model must be able to handle missing features, which naturally occur when an interaction is removed during the evaluation. Autoencoder-based models, particularly DAE, are well suited for this task because they can mask or reconstruct missing inputs without requiring a complete retraining of the model. In contrast, models such as matrix factorization (MF)~\cite{DBLP:conf/uai/RendleFGS09} or LightGCN~\cite{he2020lightgcn} must be retrained for every modification in the feature set, making them less practical for our purpose. Moreover, the DAE inherently addresses the out-of-distribution problem~\cite{vstrumbelj2014explaining,chen2023algorithms} caused by feature removal due to its robustness against noise.

Therefore, we employ the DAE as our base model, which is subsequently used to evaluate the contribution of individual items to the training loss. The DAE is composed of an encoder and a decoder. The encoder first masks the user's binary rating vector $\mathbf{r}_{u} \in \mathbb{R}^N$ to $\tilde{\mathbf{r}}_{u} \in \mathbb{R}^N$ and then converts the $\tilde{\mathbf{r}}_{u}$ into a latent representation $\mathbf{z}_{u}$, and the decoder maps $\mathbf{z}_{u}$ to the reconstruction output $\hat{\mathbf{r}}_{u}$. This process forms the basis for our Shapley value computation and ultimately allows us to determine which interactions should be considered low-quality and removed from the training data. Given the input $ \mathbf{r}_{u}$, a single hidden-layer DAE is shown as follows:
\begin{equation}
\begin{aligned}
& encoder: \mathbf{z}_{u} = a_{1}(\mathbf{W}_{1} \tilde{\mathbf{r}_{u}} + \mathbf{b}_{1}), \\
& decoder: \hat{\mathbf{r}}_{u} = a_{2}(\mathbf{W}_{2} \mathbf{z}_{u} + \mathbf{b}_{2}),
\end{aligned}
\label{encoder_and_decoder}
\end{equation}
where $\tilde{\mathbf{r}_{u}}$ is the corrupted vector of $\mathbf{r}_{u}$, $\mathbf{W}_{1} \in \mathbb{R}^{H_{1} \times N}$ and $\mathbf{W}_{2} \in \mathbb{R}^{N \times H_{1}}$ are the weight matrices, $\mathbf{b}_{1} \in \mathbb{R}^{N}$ and $\mathbf{b}_{2} \in \mathbb{R}^{H_1}$ are the bias vectors, $H_{1}$ is the dimension of the bottleneck layer, and $a_{1}$ and $a_{2}$ denote the activation functions.

To model the user preference, we follow~\cite{hu2008collaborative} to plug in a confidence matrix in the square loss function for the DAE:
\begin{equation}
\mathcal{L}_{DAE} = \sum_{u=1}^{M} \sum_{v=1}^{N} ||C_{u,v} (R_{u,v} - \hat{R}_{u,v})||_{2}^{2} = ||\mathbf{C}^{\top} \odot (\mathbf{R}^{\top} - \mathbf{\hat{R}}^{\top})||^{2}_{F},
\label{eq:AE_loss_rating}
\end{equation}
where $\odot$ is the element-wise multiplication of matrices, and $||\cdot||_{F}$ is the Frobenius norm of matrices. In particular, we set the confidence matrix $\mathbf{C} \in \mathbb{R}^{M \times N}$ as follows,
\begin{equation}
C_{u,v} =
\begin{cases}
\rho, & \text{ if } R_{u,v} = 1, \\
1, & \text{} otherwise,
\end{cases}
\label{eq:rating_weight_function}
\end{equation}
where the hyperparameter $\rho > 1$ is a constant.

So, the objective function of the overall model is as follows:
\begin{equation}
\mathcal{L} = \mathcal{L}_{DAE} + \lambda_{\mathbf{\Theta}} \Vert \mathbf{\Theta} \Vert^2,
\label{eq:final_loss}
\end{equation}
where $\lambda_{\mathbf{\Theta}}$ is the regularization parameter, and $\mathbf{\Theta}$ represents the learned parameters in the model, respectively.

\subsection{Interaction Contribution Valuation}

With the trained DAE serving as the base model, we compute Shapley values for each user--item interaction to quantify their context-aware marginal contribution to the model's training loss reduction. Formally, given an input vector $\mathbf{r} \in \mathbb{R}^{N}$ representing a user's binary interactions and the corresponding output $\mathbf{y} = f(\mathbf{r}) \in \mathbb{R}^{N}$ produced by our base recommendation model $f$, the Shapley value $\phi_i$ for the $i$-th item is defined as the average incremental change in training loss when the $i$-th interaction is added to all possible subsets of interactions.

However, the direct computation of Shapley values is combinatorially expensive, rendering it impractical for large, high-dimensional models. To overcome this challenge, we adopt FastSHAP~\cite{jethani2022fastshap}, a real-time approach that estimates Shapley values in a single forward pass. FastSHAP trains an amortized parametric function,
\begin{equation}
    \phi(\mathbf{r}, \mathbf{y}; \theta): \mathcal{R} \times \mathcal{Y} \mapsto \mathbb{R}^{N},
\end{equation}
where $\mathcal{R}$ denotes the space of input feature subsets, and $\mathcal{Y}$ represents the space of corresponding model outputs. This amortized parametric function approximates the Shapley values by minimizing a weighted least squares loss:
\begin{equation}
    \mathcal{L}(\theta) = \mathop{\mathbb{E}}\limits_{p(\mathbf{r, y})} \mathop{\mathbb{E}}\limits_{p(\mathbf{S})} \left[ \big( v_{\mathbf{r}, \mathbf{y}}(\mathbf{S}) - v_{\mathbf{r}, \mathbf{y}}(\mathbf{0}) - \mathbf{S}^{\top} \phi(\mathbf{r}, \mathbf{y}; \theta) \big)^2 \right].
    \label{eq:fastshap_opti}
\end{equation}
Here, $p(\mathbf{r, y})$ denotes the joint distribution over input-output pairs, and $p(\mathbf{S})$ is the sampling distribution over the feature subsets. The distribution $p(\mathbf{S})$ is specifically defined based on the theoretically grounded Shapley kernel~\cite{lundberg2017unified,covert2021improving} as follows,
\begin{equation}
    p(S) \propto \frac{N - 1}{\binom{N}{\mathbf{1}^\top \mathbf{S}} \cdot \mathbf{1}^\top \mathbf{S} \cdot (N - \mathbf{1}^\top \mathbf{S})}.
    \label{eqn:shapley-kernel}
\end{equation}

If the model's predictions are forced to satisfy the \textit{Efficiency Constraint}, which guarantees that the sum of Shapley values equals the total utility gain from all interactions, that is,
\begin{equation}
    \mathbf{1}^{\top} \phi(\mathbf{r}, \mathbf{y}; \theta) = v_{\mathbf{r}, \mathbf{y}}(\mathbf{1}) - v_{\mathbf{r}, \mathbf{y}}(\mathbf{0}),
\end{equation}
then given a large enough dataset and a sufficiently expressive model class for $\phi$, the global optimizer $\phi(\mathbf{r}, \mathbf{y}; \theta^*)$ is a function that outputs exact Shapley values $\phi(v_{\mathbf{r}, \mathbf{y}})$, as proved in~\cite{jethani2022fastshap}. Formally, the global optimizer satisfies the following:
\begin{equation}
    \phi(\mathbf{r}, \mathbf{y}; \theta^*)  = \phi(v_{\mathbf{r}, \mathbf{y}}) \;\; \text{almost surely in} \;\; p(\mathbf{r}, \mathbf{y}).
\end{equation}

\noindent \textbf{Efficiency Constraint Optimization}. To optimize the constrained problem in Eq.~\ref{eq:fastshap_opti} during the training process, inspired by~\cite{ruiz1998family,jethani2022fastshap}, we adopt \textit{additive efficient normalization}, which adjusts the model's outputs of $\phi(\mathbf{r}, \mathbf{y}; \theta)$ as follows:
\begin{equation}
    \label{eqn:additive-efficient-normalization}
    \phi(\mathbf{r}, \mathbf{y}; \theta) := \phi(\mathbf{r}, \mathbf{y}; \theta) + \frac{v_{\mathbf{r}, \mathbf{y}}(\mathbf{1}) - v_{\mathbf{r}, \mathbf{y}}(\mathbf{0}) - \mathbf{1}^{\top} \phi(\mathbf{r}, \mathbf{y}; \theta)}{N}.
\end{equation}

During the inference process to calculate the Shapley values, by the properties of Shapley values, non-interacted user--item pairs always yield a value of zero; therefore, our analysis focuses solely on the observed interactions $D_+$. Then, we adopt the following operation to get the final Shapley values:
\begin{equation}
    \phi(\mathbf{r}, \mathbf{y}; \theta) := \phi(\mathbf{r}, \mathbf{y}; \theta) \odot \mathbf{r}.
\end{equation}

\noindent \textbf{Value Function Design}. In our work, the value function $v_{\mathbf{r}, \mathbf{y}}(\mathbf{s})$ is designed to reflect the training loss of the DAE model. Initially, we define the value function as,
\begin{equation}
    v_{\mathbf{r}, \mathbf{y}}(\mathbf{S}) = -\frac{1}{\mathbf{1}^\top \mathbf{r}} ||f(\mathbf{r}_S) \odot \mathbf{r} -\mathbf{r}||_{2}^{2},
\end{equation}
which measures the normalized decrease in the squared error when a subset $s$ of interactions is included. Given that the implicit feedback vector $\mathbf{r}$ is binary (i.e., contains only 0s and 1s), we simplify the formulation to an equivalent form:
\begin{equation}
    v_{\mathbf{r}, \mathbf{y}}(\mathbf{S}) = \frac{\mathbf{1}^\top (f(\mathbf{r}_S) \odot \mathbf{r})}{\mathbf{1}^\top \mathbf{r}}.
\end{equation}

This equivalent form directly computes the average predicted score over the observed interactions, offering a clear, practical, and computationally efficient way to estimate the Shapley values in binary implicit feedback scenarios, where minimizing squared error loss closely aligns with maximizing predicted scores.

\subsection{Data Pruning and Retraining}
\label{sec:benchmark}

Once the Shapley values for all user-item interactions are computed, we leverage them to distinguish high-quality interactions from all interactions and retrain the base model with these selected high-quality interactions. This process involves three key steps: (1) simulated noise injection protocol, (2) contribution-driven data pruning, and (3) retraining on the pruned dataset.

\noindent \textbf{Simulated Noise Injection Protocol.} Due to the ground truth for low-quality interactions being unavailable, we employ a self-validating denoising protocol inspired by data debugging techniques~\cite{karlavs2022data,grafberger2022data}. This approach involves simulated noise generation and noise injection strategies to establish an verifiable benchmark for evaluating the effectiveness of our data pruning process. Specifically, for each user, we randomly select a fraction $k\%$ of interacted items for non-interacted items $\mathcal{N}_u$ and flip their interaction labels from $0$ to $1$, thereby treating them as artificially injected noise. Formally, the noise-injected dataset $D_{\text{noise}}$ is constructed as:
\begin{equation}
{D}_{\text{noise}} = \bigcup_{u \in \mathcal{U}} \left\{ (u, v) \mid v \sim \text{Uniform}(\mathcal{N}_u), \; |{D}_{\text{noise}}^u| = \lceil k\% \cdot |{D}_+^u| \rceil \right\},
\end{equation}
where $D_+^u$ denotes user $u$'s observed interactions. Then the perturbed training set becomes:
\begin{equation}
    {D}_{\text{corrupt}} = {D}_+ \cup {D}_{\text{noise}}.
\end{equation}

As shown in the Table~\ref{tab:noise-comparisons}, training a base recommendation model on ${D}_{\text{corrupt}}$ results in a significant drop in performance metrics (e.g., 12.52\% drop in Recall@10 and 12.43\% drop in NDCG@10 on \textit{CDs}), validating that the injected noise is indeed harmful. This provides us with a self-validating protocol to quantify the effect of injected noise on recommendation performance.

\noindent \textbf{Shapley Value-Based Data Pruning}. Given the computed Shapley values $\phi_{uv}$ for each user--item interaction, we rank all interactions based on their contributions. To filter out potential low-quality interactions for the model training, we retain the top $(1 - k)\%$ of interactions with the highest Shapley values, forming the high-quality dataset ${S}_{\text{clean}}$. The pruning process is defined as:
\begin{equation}
{S}_{\text{clean}} = \{(u, v) \in {D}_{\text{corrupt}} \mid \phi_{ui} \geq \phi_{k\%}\},
\end{equation}
where $\phi_{k\%}$ represents the Shapley value threshold corresponding to the bottom $k\%$ of ranked interactions. The removed interactions $(u, i)$ with $\phi_{ui} < \phi_{k\%}$ are regarded as likely low-quality interactions.

\noindent \textbf{Retraining with Pruned Data.} The pruned dataset ${S}_{\text{clean}}$ is subsequently used to retrain the recommendation model from scratch. Following the same procedure outlined in Sec.~\ref{sec:basemodel}, we train the base model again using the newly pruned dataset ${S}_{\text{clean}}$. Since ${S}_{\text{clean}}$ consists only of high-contribution interactions, the retrained model theoretically minimizes the negative impact of low-quality data and improves recommendation performance.

This data pruning and retraining framework ensures that the final recommendation model is trained exclusively on interactions that significantly contribute to accurate predictions, leading to enhanced robustness and reliability in recommendation performance.


\section{Experiments}
\label{sec:experiments}
In this section, we conducted experiments on four real-world datasets to demonstrate the effectiveness of our method. We mainly focus on the following questions:

\begin{itemize}[leftmargin=*]
\item \textbf{RQ1:} How does our method get better recommendation performance than the baselines?
\item \textbf{RQ2:} How effectively does our method identify injected simulated noise compared to existing approaches?
\item \textbf{RQ3:} Do the computed Shapley values align with theoretical expectations and reflect real-world interaction harmfulness?
\item \textbf{RQ4:} How robust is our method to different noise types (random, popular, or unpopular item injections)?
\item \textbf{RQ5:} Can our method provide a reasonable and intuitive case study of the calculated Shapley values?
\end{itemize}

\subsection{Experiment Setup}

\subsubsection{Datasets}

To evaluate the models under different data scales, data sparsity, and application scenarios, we performed experiments on four widely used real-world datasets. The statistics of the experimental datasets are shown in Table~\ref{dataset}.

\begin{itemize}[leftmargin=*]
    \item \textbf{Ta Feng}~\cite{hsu2004mining}: This dataset is a Chinese grocery store shopping dataset released by ACM RecSys; it covers products from food and office supplies to furniture. The dataset collected users' transaction data for 4 months, from November 2000 to February 2001.
    \item \textbf{Amazon}~\cite{ni2019justifying}: This dataset contains user reviews on products in the Amazon e-commerce system. The Amazon dataset consists of 29 sub-datasets corresponding to different product categories. For our evaluation, we choose three datasets of different categories, domains, and scales: \textit{CDs and Vinyl} (\textit{CDs}), \textit{Video Games} (\textit{Games}), and \textit{Movies and TV} (\textit{Movies}).
\end{itemize}

For all datasets, we convert all numeric ratings to implicit feedback (i.e., the user interacted with the item) and keep those with ratings as positive feedback. To ensure the quality of the dataset, we only keep the users with at least ten ratings and the items with at least twenty ratings. Specifically, for each dataset, we randomly select 80\% of historical user interactions to form the training set while treating the remaining 20\% as the test set. From the training set, we further randomly select 10\% of interactions to serve as a validation set for hyperparameter tuning. We repeat the execution of all models five times independently and report the average results.

\subsubsection{Baselines}
The main objective of this paper is to select high-quality interactions to improve the performance of recommender systems. For this purpose, we compare our model with three commonly used (implicit feedback-based) recommendation models, \textbf{BPRMF}~\cite{DBLP:conf/uai/RendleFGS09}, \textbf{LightGCN}~\cite{he2020lightgcn}, and \textbf{AE}~\cite{sedhain2015autorec,wu2016collaborative}. \textbf{DAE}~\cite{sedhain2015autorec,wu2016collaborative} serves as our base model, as introduced in Sec.~\ref{sec:basemodel}. In addition, we evaluate several data pruning strategies: \textbf{Random}, \textbf{Pred}~\cite{antwarg2021explaining}, \textbf{Sim}~\cite{xue2019deep,chalapathy2019deep}, \textbf{ADT-R}~\cite{wang2021denoising}, \textbf{ADT-T}~\cite{wang2021denoising}, and \textbf{DCF}~\cite{he2024double}. In the following, we provide detailed explanations of these methods:
\begin{itemize}[leftmargin=*]
    \item \textbf{BPRMF}: a popular method to evaluate implicit user feedback data through pairwise learning.
    \item \textbf{LightGCN}: a state-of-the-art graph model that removes the non-linear projection and embedding transformation.
    \item \textbf{AE}: reconstructs user interactions using multilayer perceptrons.
    \item \textbf{DAE}: our base model, which trains on noise-injected interactions (random masking) and reconstructs original signals.
    \item \textbf{Random}: randomly prune interactions as a naive baseline.
    \item \textbf{Pred}: prunes interactions with predicted scores closest to 0 (equivalent to high reconstruction error in autoencoders).
    \item \textbf{Sim}: prunes interactions where item similarity (e.g., cosine similarity) is lowest, assuming dissimilar interactions are noise.
    \item \textbf{ADT-R}: reweighs the loss of the user interactions based on binary cross-entropy loss, decreasing the weights of positive user-item interactions with high loss.
    \item \textbf{ADT-T}: truncates user-item interactions whose binary cross-entropy loss exceeds a dynamic threshold.
    \item \textbf{DCF}: a double correction framework for denoising recommendation via sample dropping and progressive label correction.
\end{itemize}

\begin{table}
  \caption{The statistics of datasets.}
  \label{dataset}
  \begin{tabular}{ccccc}
  \toprule
  \textbf{Dataset} & \textbf{\#User} & \textbf{\#Item} & \textbf{\#Interaction} & \textbf{Sparsity} \\
  \midrule
  \textit{Ta Feng} & 14,940 & 11,024 & 525,863 & 99.68\% \\
  \textit{CDs} & 21,667 & 39,269 & 568,489 & 99.93\% \\
  \textit{Games} & 8,255 & 17,857 & 159,854 & 99.89\% \\
  \textit{Movies} & 24,277 & 35,413 & 959,878 & 99.89\% \\
  \bottomrule
  \end{tabular}
\end{table}

\subsubsection{Evaluation Metrics}
For all experiments, we evaluate the recommendation performance in terms of accuracy (Recall@K and NDCG@K) and consider Top-$K$ ($K$ = {5, 10, 20}) lists for the recommendation. For all these metrics, the higher the value is, the better the performance is.



\begin{table*}
    \centering
    \caption{Performance comparisons of different methods. The best results are in bold, and the runners-up are with underlines in each row. R and N refer to Recall and NDCG, respectively. The Impr. represents the percentage improvement relative to the base model. Our method SVV achieves state-of-the-art results among all methods, as confirmed by a paired t-test with a significance level of 0.05.}
    \label{tab:all_comparisons}
    \resizebox{\linewidth}{!}{
    \begin{tabular}{llcccccccccccr}
    \toprule
    \textbf{Dataset} & \textbf{Metric $\uparrow$} & \textbf{BPRMF} & \textbf{LightGCN} & \textbf{AE} & \textbf{DAE} & \textbf{Random} & \textbf{Pred} & \textbf{Sim} & \textbf{ADT-R} & \textbf{ADT-T} & \textbf{DCF} & \textbf{SVV} & \textbf{Impr.} \\
    \midrule
    \multirow{6}{*}{\textit{Ta Feng}}
        & R@5 & 0.0242 & 0.0271 & 0.0295 & 0.0350 & 0.0307 & 0.0349 & 0.0348 & 0.0351 & \underline{0.0352} & 0.0348 & \textbf{0.0363} & 3.71\% \\
        & R@10 & 0.0360 & 0.0405 & 0.0431 & \underline{0.0513} & 0.0446 & 0.0509 & 0.0510 & 0.0513 & 0.0512 & 0.0511 & \textbf{0.0530} & 3.31\% \\
        & R@20 & 0.0549 & 0.0616 & 0.0634 & \underline{0.0739} & 0.0645 & 0.0732 & 0.0731 & 0.0736 & 0.0732 & 0.0730 & \textbf{0.0754} & 2.03\% \\
        & N@5 & 0.0457 & 0.0496 & 0.0531 & 0.0572 & 0.0542 & 0.0589 & 0.0587 & 0.0592 & \underline{0.0595} & 0.0584 & \textbf{0.0603} & 5.42\% \\
        & N@10 & 0.0442 & 0.0484 & 0.0513 & 0.0573 & 0.0525 & 0.0580 & 0.0580 & 0.0584 & \underline{0.0585} & 0.0579 & \textbf{0.0598} & 4.36\% \\
        & N@20 & 0.0503 & 0.0553 & 0.0579 & 0.0652 & 0.0589 & 0.0655 & 0.0654 & \underline{0.0659} & 0.0658 & 0.0653 & \textbf{0.0673} & 3.22\% \\
    \midrule
    \multirow{6}{*}{\textit{CDs}}
        & R@5 & 0.0198 & 0.0256 & 0.0429 & 0.0488 & 0.0308 & 0.0489 & 0.0487 & 0.0485 & 0.0492 & \underline{0.0497} & \textbf{0.0502} & 2.87\% \\
        & R@10 & 0.0338 & 0.0436 & 0.0643 & 0.0735 & 0.0467 & 0.0738 & \underline{0.0739} & 0.0736 & 0.0733 & 0.0735 & \textbf{0.0754} & 2.59\% \\
        & R@20 & 0.0560 & 0.0714 & 0.0926 & 0.1073 & 0.0689 & 0.1070 & 0.1071 & \underline{0.1074} & 0.1039 & 0.1045 & \textbf{0.1081} & 0.75\% \\
        & N@5 & 0.0266 & 0.0339 & 0.0560 & 0.0608 & 0.0396 & 0.0613 & 0.0612 & 0.0609 & 0.0622 & \underline{0.0629} & \textbf{0.0635} & 4.44\% \\
        & N@10 & 0.0314 & 0.0402 & 0.0628 & 0.0692 & 0.0449 & 0.0698 & 0.0699 & 0.0695 & 0.0703 & \underline{0.0708} & \textbf{0.0719} & 3.90\% \\
        & N@20 & 0.0395 & 0.0504 & 0.0731 & 0.0820 & 0.0530 & 0.0822 & 0.0823 & 0.0822 & 0.0817 & \underline{0.0824} & \textbf{0.0841} & 2.56\% \\
    \midrule
    \multirow{6}{*}{\textit{Games}}
        & R@5 & 0.0235 & 0.0289 & 0.0326 & 0.0360 & 0.0296 & 0.0376 & \underline{0.0377} & 0.0376 & 0.0374 & \underline{0.0377} & \textbf{0.0386} & 7.22\% \\
        & R@10 & 0.0394 & 0.0498 & 0.0537 & 0.0607 & 0.0482 & 0.0614 & 0.0614 & \underline{0.0618} & 0.0616 & 0.0617 & \textbf{0.0628} & 3.46\% \\
        & R@20 & 0.0638 & 0.0824 & 0.0862 & 0.0967 & 0.0780 & 0.0981 & 0.0979 & \underline{0.0986} & 0.0966 & 0.0967 & \textbf{0.0992} & 2.59\% \\
        & N@5 & 0.0266 & 0.0327 & 0.0372 & 0.0410 & 0.0337 & 0.0423 & 0.0423 & 0.0422 & 0.0422 & \underline{0.0425} & \textbf{0.0436} & 6.34\% \\
        & N@10 & 0.0330 & 0.0412 & 0.0457 & 0.0512 & 0.0412 & 0.0521 & 0.0521 & 0.0523 & 0.0522 & \underline{0.0524} & \textbf{0.0535} & 4.49\% \\
        & N@20 & 0.0424 & 0.0538 & 0.0583 & 0.0653 & 0.0527 & 0.0664 & 0.0664 & \underline{0.0666} & 0.0659 & 0.0661 & \textbf{0.0677} & 3.68\% \\
    \midrule
    \multirow{6}{*}{\textit{Movies}}
        & R@5 & 0.0168 & 0.0231 & 0.0406 & 0.0490 & 0.0313 & 0.0494 & 0.0495 & 0.0496 & 0.0496 & \underline{0.0499} & \textbf{0.0509} & 3.88\% \\
        & R@10 & 0.0285 & 0.0395 & 0.0564 & 0.0692 & 0.0450 & 0.0697 & 0.0698 & \underline{0.0704} & 0.0698 & 0.0701 & \textbf{0.0715} & 3.32\% \\
        & R@20 & 0.0479 & 0.0661 & 0.0784 & 0.0977 & 0.0653 & 0.0973 & 0.0974 & \underline{0.0983} & 0.0973 & 0.0976 & \textbf{0.0998} & 2.15\% \\
        & N@5 & 0.0302 & 0.0415 & 0.0741 & 0.0857 & 0.0569 & 0.0870 & 0.0872 & 0.0873 & 0.0877 & \underline{0.0887} & \textbf{0.0902} & 5.25\% \\
        & N@10 & 0.0322 & 0.0442 & 0.0715 & 0.0843 & 0.0558 & 0.0852 & 0.0853 & 0.0857 & 0.0857 & \underline{0.0864} & \textbf{0.0880} & 4.39\% \\
        & N@20 & 0.0396 & 0.0545 & 0.0789 & 0.0947 & 0.0631 & 0.0951 & 0.0951 & 0.0957 & 0.0954 & \underline{0.0961} & \textbf{0.0979} & 3.38\% \\
    \bottomrule
    \end{tabular}
    }
\end{table*}

\begin{table*}
    \centering
    \caption{The overlap (\%) of different methods in identifying the simulated noises. Values marked with ** indicate the overlap under an idealized setting, where injected noises perfectly correspond to harmful interactions, making higher overlap a desirable target. Bold values indicate the overlap achieved by the best-performing method in each column.}
    \begin{tabular}{crrrrrrrr}
    \toprule
    \multirow{2}{*}{\textbf{Rank}}
    & \multicolumn{2}{c}{\textit{\textbf{Ta Feng}}}
    & \multicolumn{2}{c}{\textit{\textbf{CDs}}}
    & \multicolumn{2}{c}{\textit{\textbf{Games}}}
    & \multicolumn{2}{c}{\textit{\textbf{Movies}}} \\
    \cmidrule(lr){2-3} \cmidrule(lr){4-5} \cmidrule(lr){6-7} \cmidrule(lr){8-9}
    & \multicolumn{1}{c}{\textbf{Top $\downarrow$}}
    & \multicolumn{1}{c}{\textbf{Bottom $\uparrow$}}
    & \multicolumn{1}{c}{\textbf{Top $\downarrow$}}
    & \multicolumn{1}{c}{\textbf{Bottom $\uparrow$}}
    & \multicolumn{1}{c}{\textbf{Top $\downarrow$}}
    & \multicolumn{1}{c}{\textbf{Bottom $\uparrow$}}
    & \multicolumn{1}{c}{\textbf{Top $\downarrow$}}
    & \multicolumn{1}{c}{\textbf{Bottom $\uparrow$}} \\
    \midrule
    Random & 27.00\% & 24.47\% & 24.98\% & 25.58\% & 28.15\% & 26.26\% & 22.37\% & 24.19\% \\
    Pred & 4.18\% & 53.03\% & **3.02\% & **64.61\% & 3.98\% & 60.12\% & 2.62\% & 62.40\% \\
    Sim & 21.70\% & 16.10\% & 15.82\% & 55.22\% & 23.66\% & 42.79\% & 19.26\% & 51.91\% \\
    ADT-R & 4.23\% & **53.42\% & 3.03\% & 64.24\% & **3.70\% & **60.85\% & 2.60\% & 62.13\% \\
    ADT-T & 4.19\% & 50.17\% & 3.51\% & 50.63\% & 4.00\% & 51.77\% & **2.28\% & **62.85\% \\
    DCF & **4.16\% & 50.92\% & 3.57\% & 49.41\% & 4.04\% & 51.32\% & 2.31\% & **62.85\%\\
    SVV & \textbf{6.64\%} & \textbf{41.18\%} & \textbf{11.24\%} & \textbf{47.14\%} & \textbf{10.87\%} & \textbf{47.71\%} & \textbf{8.39\%} & \textbf{45.18\%} \\
    \bottomrule
    \end{tabular}
    \label{tab:overlap}
\end{table*}

\subsubsection{Implementation Details} For a fair comparison, all methods were implemented in PyTorch and optimized using the Adam optimizer~\cite{kingma2014adam}. We maintain consistent hyperparameter settings across all methods: a latent feature dimension $H_1$ of 50, a regularization parameter $\lambda_{\mathbf{\Theta}}$ of 0.001, a learning rate of 0.001, and 200 training epochs. All parameters were initialized using Xavier initialization~\cite{glorot2010understanding}. Batch sizes were selected from \{32, 64, 128, 256\} based on the size of each dataset, and the hyperparameter $\rho$ was selected from \{5, 10, 15, 20, 25\}. For BPRMF and LightGCN, we randomly sampled 3 negative examples for each positive interaction, and LightGCN was configured with 3 graph convolutional layers. For AE and DAE, the activation function $a_1$ is \textit{ReLU}, and $a_2$ is \textit{Sigmoid}. To ensure a fair evaluation of data pruning strategies, the base model parameters and the data pruning ratio were kept constant across all methods. And the implementation of the mask binary input vector by randomly setting observed entries to zero using PyTorch's \textit{randint} function. For computational convenience, the hyperparameter $k$ was set to 20. It was found that the value of $k$ had no effect on our main experimental conclusions. When training the FastSHAP model, we employed a neural network consisting of 3 fully connected layers with 256 units per layer and \textit{ReLU} activation functions. The learning rate for FastSHAP was selected from \{0.001, 0.01, 0.1\} based on validation performance. The source code is available at \url{https://github.com/Forrest-Stone/SVV}.

\subsection{Performance Comparison (RQ1)}

Table~\ref{tab:all_comparisons} shows the performance comparisons between our approach and various baselines across multiple datasets. In summary, our experimental results validate the effectiveness of the proposed Shapley Value-driven Interaction Valuation (SVV) framework. First, SVV consistently achieves the best performance across all datasets and evaluation metrics, with an average improvement of 3.72\% over the DAE base model and an even greater improvement of 4.63\% on the Games dataset. These results demonstrate that our method more accurately identifies high-contribution interactions and mitigates the adverse effects of low-quality interactions. Second, the DAE base model outperforms traditional methods such as BPRMF, LightGCN, and AE. This finding confirms that denoising approaches are more effective than non-denoising ones and supports our choice of base model. Third, the performance of the Random baseline is substantially lower than that of the DAE, indicating that naive data pruning strategies are inadequate and more advanced approaches are required. Finally, although alternative data pruning strategies such as Pred, Sim, ADT-R, ADT-T, and DCF yield moderate improvements, their gains are marginal and sometimes even detrimental, as observed in certain metrics on the \textit{Ta Feng} dataset. In contrast, our Shapley value-based approach not only delivers significant performance improvements but also provides a transparent and theoretically grounded explanation for the importance of each interaction. These findings collectively demonstrate that SVV is a principled and effective framework for enhancing the accuracy and robustness of recommender systems by intelligently pruning low-quality data for model training while offering interpretability.

\subsection{Noise Detection Analysis (RQ2)}

Table~\ref{tab:overlap} presents the overlap of simulated noise detected by different data pruning strategies. In this table, ``Top'' refers to the proportion of noise identified among the highest-ranked interactions based on their estimated importance, whereas ``Bottom'' indicates the proportion among the lowest-ranked interactions. With $k$ set to 20 and given the division of data into training, validation, and test sets, the random baseline yields an overlap of approximately 25\%. Our analysis leads to several key conclusions. First, although none of the methods reach the theoretical optimum, our proposed SVV consistently delivers the best experimental results across all datasets and metrics. One possible reason is that the low-quality data we artificially injected may not perfectly represent the true noise labels. Second, although ADT variants and the Pred method detect a better percentage of noise in both Top and Bottom rankings, their overall recommendation performance is inferior to that achieved by SVV. This observation suggests that simply detecting more deviations from typical user behavior does not necessarily translate into higher-quality training data. Finally, the Sim method exhibits inconsistent performance, particularly on the \textit{Ta Feng} dataset, where its noise overlap is lower than that of Pred, ADT variants, and DCF, even the Random baseline. Despite this, its experimental results are close to those of Pred, ADT variants, and DCF and significantly better than Random, reinforcing the need for an efficient and interpretable data quality evaluation method.

\begin{table*}
    \centering
    \caption{Performance comparisons of different injected noise types. R and N refer to Recall and NDCG, respectively. The Impr. represents the percentage improvement relative to the base model. If the value is negative, it indicates the percentage decrease relative to the base model. The largest percentage gains are bolded in each row.}
    \label{tab:noise-comparisons}
    \begin{tabular}{ll|c|ccr|ccr|ccr}
    \toprule
    \textbf{Dataset} & \textbf{Metric $\uparrow$} & \textbf{Clean} & \textbf{Popular} & \textbf{SVV} & \textbf{Impr.} & \textbf{Random} & \textbf{SVV} & \textbf{Impr.} & \textbf{Unpopular} & \textbf{SVV} & \textbf{Impr.} \\
    \midrule
    \multirow{6}{*}{\textit{Ta Feng}}
        & R@5 & 0.0375 & 0.0325 & 0.0332 & 2.15\% & 0.0350 & 0.0363 & 3.71\% & 0.0356 & 0.0372 & \textbf{4.49\%} \\
        & R@10 & 0.0550 & 0.0479 & 0.0483 & 0.84\% & 0.0513 & 0.0530 & 3.31\% & 0.0527 & 0.0546 & \textbf{3.61\%} \\
        & R@20 & 0.0789 & 0.0685 & 0.0691 & 0.88\% & 0.0739 & 0.0754 & 2.03\% & 0.0755 & 0.0773 & \textbf{2.38\%} \\
        & N@5 & 0.0614 & 0.0556 & 0.0576 & 3.60\% & 0.0572 & 0.0603 & \textbf{5.42\%} & 0.0580 & 0.0610 & 5.17\% \\
        & N@10 & 0.0613 & 0.0548 & 0.0561 & 2.37\% & 0.0573 & 0.0598 & 4.36\% & 0.0583 & 0.0610 & \textbf{4.63\%} \\
        & N@20 & 0.0695 & 0.0615 & 0.0628 & 2.11\% & 0.0652 & 0.0673 & 3.22\% & 0.0663 & 0.0687 & \textbf{3.62\%} \\
    \midrule
    \multirow{6}{*}{\textit{CDs}}
        & R@5 & 0.0545 & 0.0422 & 0.0419 & -0.71\% & 0.0488 & 0.0502 & 2.87\% & 0.0514 & 0.0535 & \textbf{4.09\%} \\
        & R@10 & 0.0827 & 0.0632 & 0.0623 & -1.42\% & 0.0735 & 0.0754 & 2.59\% & 0.0772 & 0.0802 & \textbf{3.89\%} \\
        & R@20 & 0.1180 & 0.0922 & 0.0886 & -3.90\% & 0.1073 & 0.1081 & 0.75\% & 0.1108 & 0.1157 & \textbf{4.42\%} \\
        & N@5 & 0.0681 & 0.0532 & 0.0540 & 1.50\% & 0.0608 & 0.0635 & 4.44\% & 0.0640 & 0.0671 & \textbf{4.84\%} \\
        & N@10 & 0.0778 & 0.0601 & 0.0605 & 0.67\% & 0.0692 & 0.0719 & 3.90\% & 0.0728 & 0.0761 & \textbf{4.53\%} \\
        & N@20 & 0.0910 & 0.0708 & 0.0701 & -0.99\% & 0.0820 & 0.0841 & 2.56\% & 0.0853 & 0.0893 & \textbf{4.69\%} \\
    \midrule
    \multirow{6}{*}{\textit{Games}}
        & R@5 & 0.0413 & 0.0333 & 0.0324 & -2.70\% & 0.0360 & 0.0386 & 7.22\% & 0.0381 & 0.0411 & \textbf{7.87\%} \\
        & R@10 & 0.0678 & 0.0539 & 0.0532 & -1.30\% & 0.0607 & 0.0628 & 3.46\% & 0.0635 & 0.0672 & \textbf{5.83\%} \\
        & R@20 & 0.1068 & 0.0868 & 0.0846 & -2.53\% & 0.0967 & 0.0992 & 2.59\% & 0.1021 & 0.1051 & \textbf{2.94\%} \\
        & N@5 & 0.0463 & 0.0373 & 0.0370 & -0.80\% & 0.0410 & 0.0436 & 6.34\% & 0.0428 & 0.0461 & \textbf{7.71\%} \\
        & N@10 & 0.0573 & 0.0458 & 0.0454 & -0.87\% & 0.0512 & 0.0535 & 4.49\% & 0.0534 & 0.0570 & \textbf{6.74\%} \\
        & N@20 & 0.0725 & 0.0586 & 0.0576 & -1.71\% & 0.0653 & 0.0677 & 3.68\% & 0.0684 & 0.0717 & \textbf{4.82\%} \\
    \midrule
    \multirow{6}{*}{\textit{Movies}}
        & R@5 & 0.0520 & 0.0465 & 0.0473 & 1.72\% & 0.0490 & 0.0509 & \textbf{3.88\%} & 0.0506 & 0.0525 & 3.75\% \\
        & R@10 & 0.0738 & 0.0661 & 0.0665 & 0.61\% & 0.0692 & 0.0715 & 3.32\% & 0.0719 & 0.0743 & \textbf{3.34\%} \\
        & R@20 & 0.1035 & 0.0924 & 0.0925 & 0.11\% & 0.0977 & 0.0998 & 2.15\% & 0.1004 & 0.1039 & \textbf{3.49\%} \\
        & N@5 & 0.0912 & 0.0811 & 0.0840 & 3.58\% & 0.0857 & 0.0902 & \textbf{5.25\%} & 0.0891 & 0.0924 & 3.70\% \\
        & N@10 & 0.0896 & 0.0798 & 0.0819 & 2.63\% & 0.0843 & 0.0880 & \textbf{4.39\%} & 0.0875 & 0.0906 & 3.54\% \\
        & N@20 & 0.1002 & 0.0893 & 0.0909 & 1.79\% & 0.0947 & 0.0979 & 3.38\% & 0.0977 & 0.1011 & \textbf{3.48\%} \\
    \bottomrule
    \end{tabular}
\end{table*}

\begin{table}
    \centering
    \caption{The overlap (\%) of different injected noise types for identifying the simulated noises. The best results are in bold in each row.}
    \begin{tabular}{lrrrr}
    \toprule
        \textbf{Dataset} & \textbf{Rank} & \textbf{Popular} & \textbf{Random} & \textbf{Unpopular} \\
        \midrule
        \multirow{2}{*}{\textit{Ta Feng}} & Top $\downarrow$ & 16.48\% & 6.64\% & \textbf{1.94\%} \\
         & Bottom $\uparrow$ & 20.92\% & 41.18\% & \textbf{41.30\%} \\
        \midrule
        \multirow{2}{*}{\textit{CDs}} & Top $\downarrow$ & 21.22\% & 11.24\% & \textbf{1.22\%} \\
         & Bottom $\uparrow$ &27.54\% & 47.14\% & \textbf{64.70\%} \\
        \midrule
        \multirow{2}{*}{\textit{Games}} & Top $\downarrow$ & 21.86\% & 10.87\% & \textbf{2.76\%} \\
         & Bottom $\uparrow$ & 27.30\% & 47.71\% & \textbf{58.53\%} \\
        \midrule
        \multirow{2}{*}{\textit{Movies}} & Top $\downarrow$ & 12.58\%  & 8.39\% & \textbf{0.89\%} \\
         & Bottom $\uparrow$ & 36.98\% & 45.18\% & \textbf{60.46\%} \\
    \bottomrule
    \end{tabular}
    \label{tab:noise-overlap}
\end{table}

\subsection{Shapley Value Validation (RQ3)}

\begin{figure}
    \centering
    \captionsetup[subfigure]{aboveskip=7pt,belowskip=7pt}
    \subcaptionbox{Segment exclusion of interactions ranked by Shapley values (left: low-to-high order; right: high-to-low order). \label{fig:a}}
    {
        \includegraphics[width =0.49\linewidth]{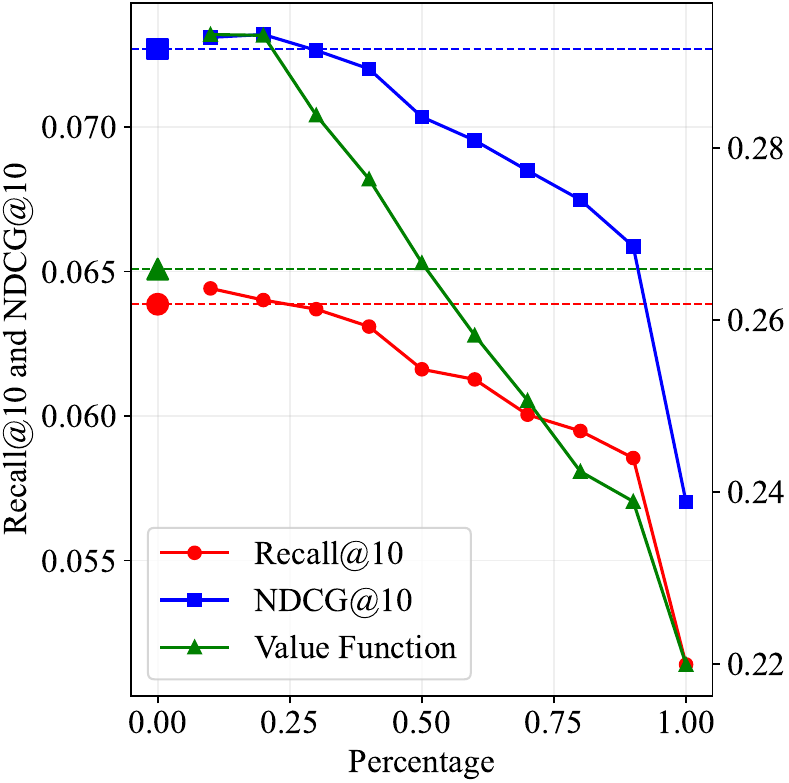}
        \includegraphics[width =0.49\linewidth]{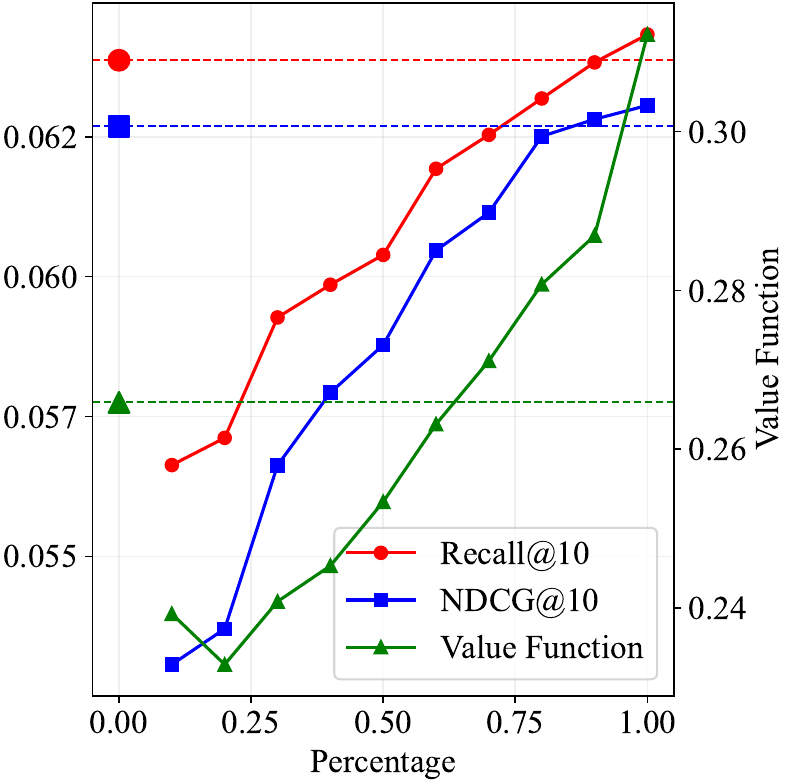}
    }
    \subcaptionbox{Cumulative exclusion of interactions ranked by Shapley values (left: low-to-high order; right: high-to-low order). \label{fig:b}}
    {
        \includegraphics[width =0.49\linewidth]{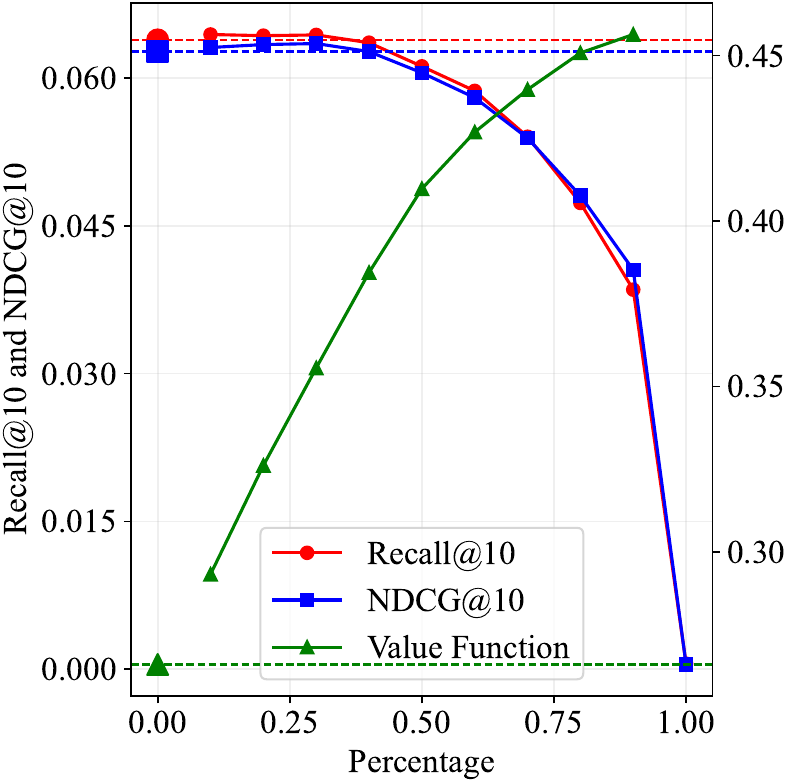}
        \includegraphics[width =0.49\linewidth]{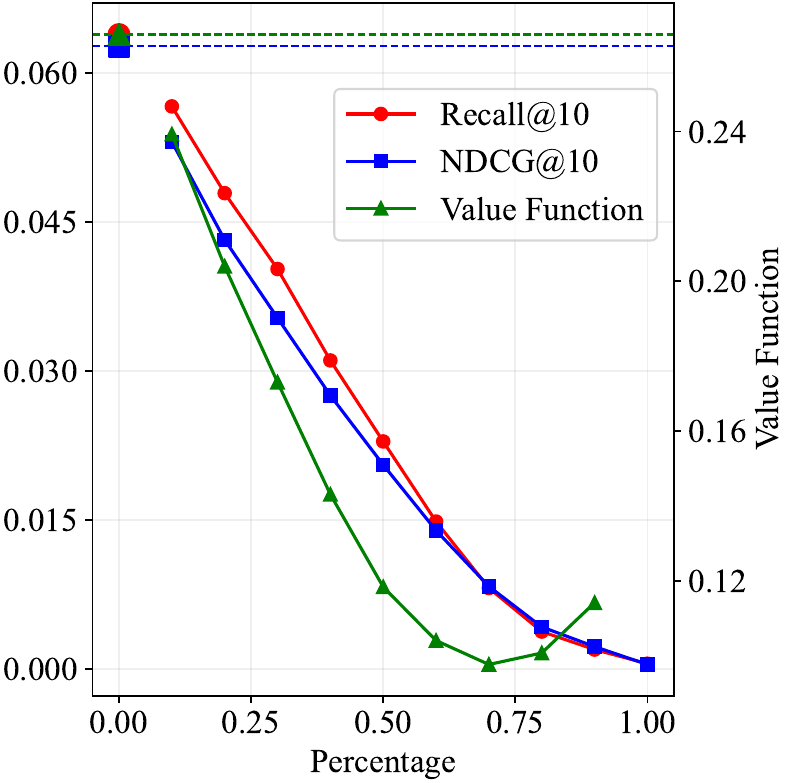}
    }
    \setlength{\abovecaptionskip}{4pt}
    \caption{Contrasting Shapley value-based exclusion strategies: segment vs. cumulative impact on recommendation performance (Recall@10 and NDCG@10) and model training (Value Function) on \textit{CDs} dataset.}
    \Description{Contrasting Shapley value-based exclusion strategies: segment vs. cumulative impact on recommendation performance (Recall@10 and NDCG@10) and model training (Value Function) on \textit{CDs} dataset.}
    \label{fig:exclusion}
\end{figure}

Figure~\ref{fig:exclusion} compares two exclusion strategies based on Shapley values, segmented and cumulative, and their effects on recommendation performance (Recall@10 and NDCG@10) and model training (as measured by the value function). The results of the segmented exclusion approach are shown in Figure~\ref{fig:a}. When interactions are removed in segments in ascending order by their Shapley values, the performance metrics and the value function steadily decline; the reverse ordering yields the opposite trend. These trends indicate that the calculated Shapley value in our method is accurate, and interactions with lower Shapley values adversely affect training and can be considered low-quality data. Figure~\ref{fig:b} presents the results of the cumulative exclusion strategy. When interactions are removed in ascending order, performance metrics remain stable initially and then decline as the value function gradually increases. In contrast, when interactions are removed in descending order, the value function decreases at first and then slightly rebounds, while performance metrics continuously decline. These observations validate the effectiveness of our simulated noise injection and underscore the importance of accurately evaluating training data quality. Overall, these results reinforce our core idea that an evaluation based on Shapley values provides an efficient and interpretable means to identify and prune low-quality interactions, thereby enhancing recommendation accuracy and robustness.

\begin{table*}
    \centering
    \caption{A case study of SVV's interpretability on user \textit{Super Seven}.}
    \label{tab:case-study}
    \begin{tabular}{llrl}
        \toprule
        \textbf{Item ID} & \textbf{Title} & \textbf{Shapley Value} & \textbf{Category} \\
        \midrule
        $v_{35251}$ & Sunbathing Animal & -0.00253 & Alternative Rock \\
        $v_{5871}*$ & Tribute to Woody Guthrie & 0.00516 & Classic Rock, Album-Oriented Rock (AOR) \\
        $v_{5283}$ & Fountains Of Wayne & 0.00595 & Alternative Rock, Indie and Lo-Fi, Indie Rock \\
        $v_{39187}$ & The Complete Studio Albums & 0.00927 & Alternative Rock \\
        $v_{17966}*$ & Madrigals & 0.01059 & Alternative Rock, Singer-Songwriters \\
        $v_{1846}$ & John Barleycorn Must Die & 0.01375 & Rock, Progressive, Progressive Rock \\
        $v_{1780}$ & Welcome to the Canteen & 0.01518 & Rock, Progressive, Progressive Rock \\
        $v_{14732}*$ & Legacy & 0.01527 & Alternative Rock, Indie and Lo-Fi, Indie Rock \\
        $v_{25}$ & Retrospective Hybrid & 0.01531 & Pop \\
        $v_{36524}$ & Still the King & 0.01677 & Country, Classic Country \\
        $v_{501}$ & Shoot Out at the Fantasy Factory & 0.01729 & Rock, Progressive, Progressive Rock \\
        $v_{32918}$ & Same Trailer Different Park & 0.03358 & Country, Today's Country \\
        \bottomrule
    \end{tabular}
\end{table*}

\subsection{Noise Type Robustness (RQ4)}

Table~\ref{tab:noise-comparisons} and Table~\ref{tab:noise-overlap}  present the impact of different noise types on our model and the overlap analysis for detecting simulated noise. In our experiments, items with the top 20\% of user interactions seem popular, while those with the bottom 20\% seem unpopular. ``Clean'' represents no added noise, ``Random'' refers to randomly added noise, ``Popular'' indicates the injection of popular items, and ``Unpopular'' represents the addition of unpopular items. As observed in Table~\ref{tab:noise-comparisons}, when low-quality data from unpopular items is added, our method significantly improves the base model's performance, particularly on the \textit{Movies} dataset, even surpassing the Clean condition. However, adding popular items results in a deterioration of performance, likely because popular items distort the true user preferences, masking their actual behavior. This implies that careful consideration of how data influences model training is essential. Table~\ref{tab:noise-overlap} further reveals that our method is most effective at detecting noise from unpopular items, which is optimal across all datasets. These findings suggest that our method relies on a robust base model and is well-suited for handling long-tail noise, reinforcing the importance of accurately identifying and managing noise in recommender systems.

\subsection{Case Study (RQ5)}

To demonstrate the interpretability of SVV, we present a case study on user \textit{Super Seven} from the CDs dataset, as shown in Table~\ref{tab:case-study}. We rank this user's interactions by Shapley values and identify the bottom three items (shown in the first three rows of the table) as low-quality data for pruning. Items marked with * are injected noises, and we observe they do not fully overlap with the lowest-ranked interactions, which supports our motivation that user intent does not always reflect training utility. For example, Sunbathing Animal (Item ID: $v_{35251}$) received a high rating of 5 but was assigned a low Shapley value, indicating minimal contribution to model learning. This highlights that SVV can distinguish user-preferred items from those truly valuable for training, revealing that some positive feedback may have little learning utility, offering transparency that black-box or intent-based methods typically lack.

\section{Conclusion and Future Work}
In this paper, we present Shapley Value-driven Valuation (SVV), a principled framework for data pruning in recommender systems that shifts the focus from subjective intent-based denoising to an objective, model-driven assessment of training utility. SVV quantifies each user--item interaction's Shapley value by measuring its marginal contribution to reducing training loss, efficiently computed using FastSHAP.
Moreover, by incorporating a simulated noise injection protocol, we establish a verifiable benchmark that objectively validates the impact of noise on model performance. Extensive experiments on four real-world datasets demonstrate that SVV outperforms traditional denoising methods in terms of accuracy and robustness by preserving interactions that are critical for effective model training. Future work will extend SVV to dynamic settings and explore scalable variants to further enhance its applicability in evolving recommendation environments.

\begin{acks}
This work was supported by the Early Career Scheme (No. CityU 21219323) and the General Research Fund (No. CityU 11220324) of the University Grants Committee (UGC), and the NSFC Young Scientists Fund (No. 9240127). 
\end{acks}

\bibliographystyle{ACM-Reference-Format.bst}
\balance
\bibliography{06_references.bib}


\end{document}